\documentstyle[aps,amstex,epsfig,float]{revtex}
\epsfverbosetrue
\newcommand{\beq}{\begin{equation}}
\newcommand{\eeq}{\end{equation}}
\newcommand{\beqn}{\begin{eqnarray}}
\newcommand{\eeqn}{\end{eqnarray}}
\newcommand{\beqns}{\begin{eqnarray*}}
\newcommand{\eeqns}{\end{eqnarray*}}

\newcommand{\intl}{\int\limits}

\def\NPs{{\it Nucl. Phys. (Proc. Suppl.)}}
\def\PL{{\it Phys. Lett.}}
\def\PR{{\it Phys. Rev.}}

\def\PRL{{\it Phys. Rev. Lett.}}
\def\NIM{{\it Nucl. Inst. Meth.}}

\def\EPJ{{\it Europ. Phys. J.}}

\def\NC{{\it Nuov. Cim.}}
\def\AP{{\it Astropart. Phys.}}
\def\Na{{\it Nature}}
\def\ApJ{{\it Astrophys. J.}}
\def\RMP{{\it Rev. Mod. Phys.}}
\def\AA{{\it Astron. Astrophys.}}
\def\ea{{\it et al.}}
\begin{document}


\title{
  Gravity Wave and Neutrino Bursts from Stellar Collapse:
    A Sensitive Test of Neutrino Masses
}

\author{
N.~Arnaud, M.~Barsuglia, M.~A.~Bizouard, F.~Cavalier, M.~Davier,
P.~Hello and T.~Pradier
}

\address{
Laboratoire de l'Acc\'el\'erateur Lin\'eaire,
IN2P3-CNRS et Universit\'e de Paris-Sud, F-91898 Orsay, France\protect\\
}

\maketitle
\begin{abstract}
New methods are proposed with the goal to determine 
absolute neutrino masses from the simultaneous observation 
of the bursts of neutrinos and gravitational waves 
emitted during a stellar collapse. It is shown that the
neutronization electron neutrino flash and the 
maximum amplitude of the gravitational wave signal
are tightly synchronized with the bounce occuring at the end of
the core collapse on a timescale better than 1 ms. The existing 
underground neutrino detectors (SuperKamiokande, SNO, ...) and
the gravity wave antennas soon to operate (LIGO, Virgo, ...) are well 
matched in their performance for detecting galactic supernovae and 
for making use of the proposed approach. Several methods are described,
which apply to the different scenarios depending on neutrino
mixing. Given the present knowledge on neutrino oscillations, the
methods proposed are sensitive to a mass range where neutrinos 
would essentially be mass-degenerate.
The $95~\%$ C.L. upper limit which can be achieved varies 
from 0.75~eV/c$^2$ for large $\nu_e$ survival probabilities 
to 1.1~eV/c$^2$ when in practice all $\nu_e$'s convert 
into $\nu_\mu$'s or $\nu_\tau$'s. The sensitivity is nearly independent 
of the supernova distance. 
\end{abstract}


\baselineskip = 2\baselineskip

%
%
\section{Introduction }
The understanding of the origin of the tiny neutrino mass scale is one of 
the most puzzling problems in fundamental physics. On one hand finite
neutrino masses of order 1~eV/c$^2$ or below indicate new physics 
beyond the Standard Model, as such masses are generally 
induced from a large mass scale (see {\it e.g.} Ref.~\cite{seesaw}), 
possibly as large as the Planck scale. On the other hand 
neutrino masses of order 1~eV/c$^2$ have
cosmological implications as relic neutrinos could represent a
significant part of dark matter. Recent analyses 
of galaxy clustering in the context of a nonzero
cosmological constant tend to limit the contribution of hot dark
matter (neutrinos) to masses less than 4~eV/c$^2$~\cite{gawiser}.
The limit can be lowered to 2.2~eV/c$^2$ when the recent data on CMB 
anisotropies are included~\cite{wang}. 

Strong experimental evidence has been recently presented for neutrino
flavour oscillations~\cite{osakanu}. Although the complete
picture is not totally clear, the most solid interpretation of the
reduced solar $\nu_e$ flux on Earth (see for instance Ref.~\cite{krastev}
for a recent analysis) and the $\nu_\mu$ deficit in
atmospheric production by cosmic rays as detected by underground
experiments relies on mixing, where the mass eigenstates $\nu_i$ are 
linear combinations of the 3 neutrino flavour states. The current scenario
is based on ({\it i}) $\nu_e-\nu_\mu$ oscillations with four distinct 
solutions, three with near-maximal mixing and $\Delta m_{12}^2 = |m_{\nu_1}^2 
- m_{\nu_2}^2|  \sim 10^{-10}$, $\sim 10^{-7}$  
or $\sim 5~10^{-5}$ eV$^2$/c$^4$ and the fourth (less favoured) with
small mixing and $\Delta m_{12}^2  \sim 10^{-5}$~eV$^2$/c$^4$, and ({\it ii})
$\nu_\mu-\nu_\tau$ oscillations with a unique solution characterized
by maximal mixing and  $\Delta m_{23}^2  \sim 3.5~10^{-3}$~eV$^2$/c$^4$.
Several experimental programs are underway in order to confirm the
interpretation in terms of oscillations~\cite{osakanu}.

Even if nonzero $\Delta m_{ij}^2$ are nearly established the absolute 
neutrino mass scale is still unknown. Direct measurements of neutrino
masses provide us only with upper limits~\cite{pdg2000}: 
3~eV/c$^2$ for $\nu_e$~\cite{mainz,troitsk}, 190~keV/c$^2$ for $\nu_\mu$ 
and 18~MeV/c$^2$ for $\nu_\tau$. In the context of the
neutrino oscillations discussed above only the $\nu_e$ mass limit is 
relevant. 
Putting together this limit and the oscillation results, 
two extreme scenarios for the neutrino mass spectrum
can be considered: ({\it i}) a
spreadout spectrum with $m_{\nu_\tau} \sim 60$~meV/c$^2$, 
$m_{\nu_\mu} \sim 3$~meV/c$^2$ and $m_{\nu_e} \ll 3$~meV/c$^2$, 
or ({\it ii}) a nearly degenerate spectrum
with a common mass as large as 3~eV/c$^2$ and a splitting determined by the
small $\Delta m_{ij}^2$ from the observed oscillations. While the first 
solution looks more natural, {\it i.e.} resembling the charged lepton and
quark mass pattern, the second one is cosmologically 
more interesting and might also be easier to understand 
in the context of maximal mixing, a feature 
quite different from what is observed with quarks. 

It is therefore very important to investigate the possibility to 
directly measure neutrino masses below the current $\nu_e$ mass limit.
In this paper a new method is proposed to determine neutrino masses
by exploiting the timing between the bursts of gravitational waves
(GW) and of neutrinos emitted just at the end of the collapsing phase of a 
supernova. This technique capitalizes on the availability of operating
underground detectors which are well suited to the neutrino energy
range from supernovae (SuperKamiokande~\cite{superk}, SNO~\cite{sno},
and other less sensitive detectors~\cite{lvd,macro}) and on forthcoming GW
interferometric antennas (LIGO~\cite{ligo}, Virgo~\cite{virgo}, and 
others~\cite{geo,tama}) whose sensitivity to short bursts is well matched.
The combination of an astronomical baseline and millisecond timing 
allows one to reach the 1~eV/c$^2$ level for the neutrino mass 
and possibly better. 
Several variants are proposed which match the performances of the neutrino 
detectors and apply in different scenarios for neutrino mixing.

Many studies have already been performed on the possibility to use 
supernova explosions to measure or bound neutrino masses. Following 
the first neutrino observations from SN1987A~\cite{kamioka,imb} 
$\nu_e$ mass limits have been obtained around 20~eV/c$^2$~\cite{loredo} 
using the time spread of the burst a few seconds long which would be 
sensitive to massive neutrinos. Other methods have been proposed for 
the next near-galactic supernova occurence~\cite{beacom1,beacom2} with 
sensitivities reaching 3~eV/c$^2$~\cite{totani}. In the case where the 
stellar core collapses early into a black hole, the neutrino production
is suddenly quenched, providing a method with an estimated mass 
sensitivity of 1.8~eV/c$^2$~\cite{beacom3}.

%
%
\section{Supernova Dynamics}
The physics of Type II stellar collapse and the subsequent radiation of 
gravitational waves and of neutrinos has been a subject of intense 
research for more than 40 years. Extensive reviews and references can be 
found~\cite{bethe,burrows}. Here we only recall the main
model-independent features on which our approach is based.

The infalling iron core of the star produces electron neutrinos
when electrons are captured by protons. The core collapse is homologous 
and as nuclear densities are exceeded it becomes opaque to neutrinos 
which are captured inside. The small nuclear compressibility brings 
the collapse to a halt, producing a bounce which generates a strong 
shock wave travelling back through the neutrinosphere, at which 
point the medium becomes transparent enough for the neutrinos 
to escape. This generates a short $\nu_e$ flash and
signals the onset of the emission of all neutrino types produced
thermally as $\nu \bar{\nu}$ pairs from the heat generated on
the accretion surface during infall. Unlike for the $\nu_e$ burst the
thermal emission is expected to last a few seconds. The main point is 
the strong time correlation between the bounce and the $\nu_e$ 
flash generated by neutronization in the low-density
outer part of the core. The flash delay and its duration are controlled by
the shock dynamics whose description is expected to be strongly 
model-dependent.
However, the timescale involved is so short that it can be determined
on quite general ground by hydrodynamics considerations
~\cite{bethe_ab,mazurek,lattimer,burrows_kg}.

The shock wave is generated deep into the core ($ r \sim 10$ km) and
propagates outward with a velocity $v \sim 0.1~c$ whose precise value
depends on the  shock strength. The shock reaches the 
neutrinosphere at a radius $r_0 \sim 90 {\rm km}$
defined such that the neutrinos see only one absorption length of matter 
outside of it. The number of $\nu_e$'s rises fastly and then decays
exponentially. Many estimates have been given in the 
literature~\cite{mazurek}, the most recent ones from sophisticated 
hydrodynamical simulations~\cite{simul}. The mean timing of the 
$\nu_e$ pulse with respect to the bounce turns out to be
   $\Delta t_{\nu_e,bounce} = (3.5 \pm 0.5) ~{\rm ms}$~.
While the above estimates of $v$, $r_0$ and subsequently 
$\Delta t_{\nu_e,bounce}$ depend on the properties of the compressed
nuclear matter, valuable information can be deduced from observables
such as the mean $\nu_e$ energy in the flash~\cite{mazurek}, thus 
helping through simulation to constrain the range of the relevant 
parameters. Finally, the integrated luminosity in the $\nu_e$ flash is 
estimated to be $\intl L_{\nu_e}~dt \sim 3~10^{51}~$ erg, corresponding 
to about $1~\%$ of the total energy carried away by neutrinos in the  
few seconds following the initial collapse.

\begin{figure}
\begin{center}
\epsfxsize 8cm
\epsffile{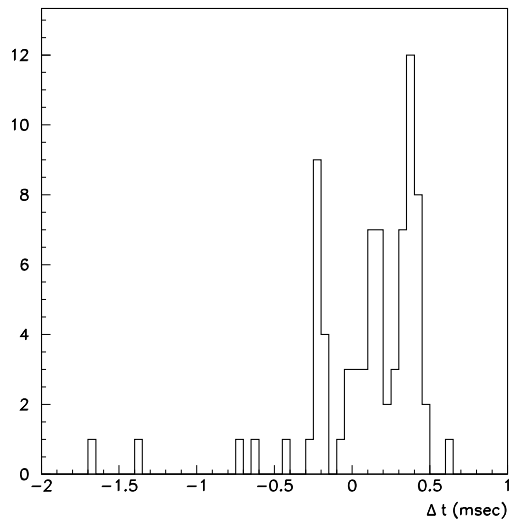}
\end{center}
\vspace{-0.5cm}
\caption{\it Distribution of the time difference between the maximum of the
          GW amplitude and the core collapse bounce for the signals
            simulated in Ref.~[34].}
\label{zm_tbmax}
\end{figure}

The fast core collapse and the resulting bounce are expected to produce
radiation of gravitational waves. Many hydrodynamical simulations assuming
specific core models have been performed~\cite{sn_gw}. It is observed that
the details of the produced waveform are highly model-dependent. In
particular the rotation of the inner core is found to be an important factor 
as centrifugal forces tend to delay the collapse or even sometimes prevent it
altogether. At any rate a strong correlation in time is expected between
the core bounce and the maximum of gravitational radiation. This effect
has been studied with specific collapse models. For example we used the
library of 78 typical waveforms which has been produced in 
simulations~\cite{zm}, varying rotation and equation-of-state parameters
within reasonable ranges. Despite a strong variability in the signal shape,
the location of the maximum wave amplitude is tightly correlated to the
bounce as shown in Fig.~\ref{zm_tbmax}. In fact the signals with 'abnormal' 
delays are most of the time characterized by a relatively smaller amplitude
and are therefore less likely to be detected in the first place 
by the GW interferometers.

Thus our method is based on the strong time correlation between the 
$\nu_e$ flash and the peak of gravitational radiation in the event of
a type II supernova.

%
%
\section{Neutrino Detection}
\label{nudetect}
Detection of the $\nu_e$ flash is possible with already operating experiments.
The most sensitive ones are SuperKamiokande~\cite{superk} and SNO~\cite{sno}
which are both large volume water Cerenkov detectors.

SuperKamiokande can detect $\nu_e$'s, as well as all neutrino types, through 
elastic scattering on electrons
\beqn\label{nue}
      \nu_i ~e^- ~\longrightarrow ~\nu_i ~e^-~.
\eeqn
While this process has the advantage of being directional 
---essentially all events are concentrated in a cone $\cos{\theta}>0.8$,
where $\theta$ is the angle between the source direction and the electron 
recoil--- it suffers from the fact that information on the incident neutrino
energy $E_\nu$ is lost, the electron energy spectrum being uniformly 
distributed between 0 and $E_\nu$. Thermal neutrinos such as 
$\overline{\nu}_e$'s can be detected through the charged-current process
\beqn\label{antinup}
      \overline{\nu}_e ~p ~\longrightarrow ~e^+ ~n~.
\eeqn
This reaction is essentially isotropic, therefore carrying no information on
the source direction, but it allows a direct measurement of 
$E_\nu = E_e + E_{th}$, with the threshold energy $E_{th} \simeq 1.77$ MeV.

As a heavy water Cerenkov detector, SNO has unique capabilities 
for detecting $\nu_e$'s and $\overline{\nu}_e$'s by means of the charged 
current processes on deuterons
\beqn\label{nudcc1}
      \nu_e ~d ~  \longrightarrow  ~e^- ~p ~p~, 
\eeqn
\beqn\label{nudcc2}
      \overline{\nu}_e ~d ~  \longrightarrow  ~e^+ ~n ~n~,
\eeqn
and all neutrino types through the neutral-current reaction
\beqn\label{nudnc}
      \nu_i ~d ~  \longrightarrow  ~\nu_i ~p ~n~.
\eeqn
All reactions are isotropic, with energy measurement for the charged
current processes, with $E_{th} \simeq 1.44$ MeV for $\nu_e$ and 4.03 MeV 
for $\overline{\nu}_e$. The neutral-current processes are detected
using neutron capture by $^{35}{\rm Cl}$ in dissolved salt, leading to an
8.6 MeV $\gamma$ ray. While all the reactions discussed so far have
excellent timing, of the order of a few tens of ns, the situation is not 
as good for the neutral-current ones where the detection timing
is limited by neutron diffusion, inducing an exponentially distributed delay
with a time constant of $\sim$4 ms~\cite{sno}.

The event rates are large enough for galactic supernovae. 
Relevant cross sections and their energy dependence can be found 
in Ref.~\cite{burrows_kg}. Using the luminosity given above for a 
supernova exploding 10 kpc away, the expected numbers are 15 for 
the $\nu_e$ flash and 5300 for thermal $\overline{\nu}_e$'s through 
processes (\ref{nue}) and (\ref{antinup}) respectively, in SuperKamiokande.
Similarly, 13 events are expected for the $\nu_e$ flash through 
process~(\ref{nudcc1}) in SNO. All these rates scale as $L^{-2}$, where
$L$ is the supernova distance.

%
%
\section{Effect of Neutrino Mixing}
The propagation of neutrinos from the star to the detectors can be
affected by flavour oscillations. Starting with the observation of 
neutrinos from SN1987A this question has been studied by many 
authors, considering both vacuum and matter-enhanced 
oscillations~\cite{nuoscil}. For our purpose 
it is important on one hand to estimate the $\nu_e$ survival 
probability, $P_e$, affecting the total rate for charged-current processes 
and consequently the statistical power of the measurement. 
On the other hand if $P_e$ gets too small the 
neutronization flash will arrive on Earth mostly as $\nu_\mu$'s or 
$\nu_\tau$'s which can only be detected by neutral-current reactions. The SNO
detector is well suited to this purpose, with however a worsening of the 
timing resolution due to fluctuations in the neutron capture, 
as discussed above.
 
A comprehensive treatment of oscillations for neutrinos born in a stellar
collapse has been recently presented~\cite{dighe} and we follow here this
analysis. The MSW resonances~\cite{msw} play a crucial role while the
neutrinos propagate in the matter of slowly-decreasing density. The effect
on $\nu_e$'s is in general important, but depends crucially on the solar
neutrino oscillation solution and whether the neutrino mass hierarchy is
'normal' (the mass eigenstates $\nu_1$, $\nu_2$ and $\nu_3$ have increasing
masses) or 'inverted' (the $\nu_1$ mass state, mostly connected to the
$\nu_e$ flavour state, is the heaviest). We remark that if the mass states
are spread out, as for the other fermions, it is more natural to expect the
'normal' hierarchy, while for a quasi-degenerate spectrum both scenarios are
equally plausible. It turns out that the value of $P_e$ depends in a strong
way on the mixing matrix element $U_{e3}$ between $\nu_e$ and $\nu_3$ states,
the other relevant elements being fixed by unitarity and the solar mixing 
angle. The only known experimental constraint on $U_{e3}$ comes 
from the Chooz reactor oscillation experiment~\cite{chooz}, 
yielding $|U_{e3}|^2 < 3~10^{-2}$.

\begin{figure}
\begin{center}
\epsfxsize 8cm
\epsffile{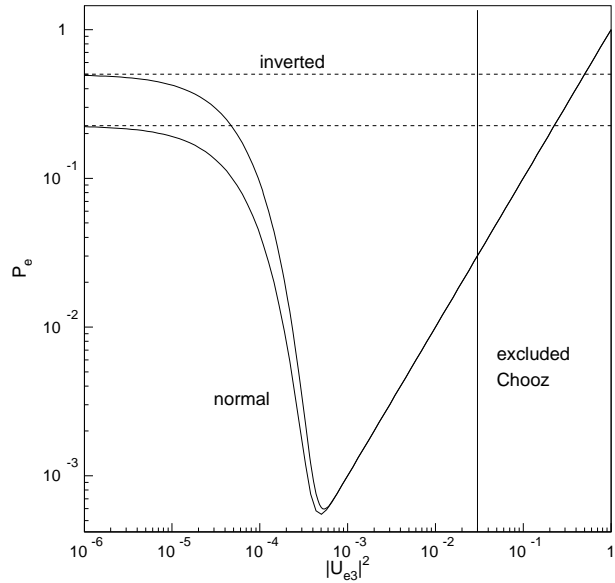}
\end{center}
\caption{\it The electron survival probability from a stellar collapse to
             Earth in the large mixing-angle MSW scenario for solar 
             neutrinos as a function of the matrix element $|U_{e3}|^2$.
             The solid curves correspond to the 'normal' mass
             hierarchy, while the dashed ones stand for the 'inverted'
             hierarchy. In each case the upper (lower) curve is computed with
             $\sin{2 \theta_\odot}^2 = 1.0~(0.7)$. The calculations follow
             the analysis given in Ref.~[36]. The area corresponding 
             to $|U_{e3}|^2 > 3~10^{-2}$ is excluded by the Chooz 
             experiment~[38].}
\label{oscil}
\end{figure}

Fig.~\ref{oscil} shows the situation for the large mixing-angle 
($\theta_\odot$) MSW solution, with $\sin{2 \theta_\odot}^2 = 0.7-1.0$,
which seems to be favoured by experimental data~\cite{osakanu}. In this case
the $\nu_e$ peak is still preserved with a survival probability between 0.2 
and 0.5, except for values of $|U_{e3}|^2 > 10^{-5}$ in the normal
hierarchy scenario. The small mixing-angle MSW and the vacuum oscillation
solutions yield different behaviours with $P_e$ values ranging from 0.8 to
negligible. In the following we shall use a value $P_e = 0.5$ as 
representive of situations where the $\nu_e$ flash content is well 
preserved, and also consider the case where the $\nu_e$ rate becomes 
too small rendering $\nu_{\mu,\tau}$ detection mandatory. In this way 
all possibilities are covered. It should be remarked that, contrary 
to the $\nu_e$ case, the rate of $\overline{\nu}_e$'s remains 
essentially unaltered, to the extent that thermal production should
result in approximately equal numbers of neutrino pairs of each flavour.

%
%
\section{Relative Timing}
\label{timing}
We are now in position to discuss the relative timing of the neutrino and
gravitational wave bursts. Both emission times have been seen to be closely 
related to the bounce time in the core collapse. 

Travel times to Earth depend on neutrino and graviton masses. 
Very constraining bounds exist on the graviton mass $m_G$: 
in particular, precise studies of planet orbits in
the solar system~\cite{will} yield a lower limit for the graviton Compton
wavelength $\lambda_G = h / m_G c$ of $3~10^{12}$ km, much larger
than the value $6~10^{9}$ km that would produce a time delay equal to
that of a 1 eV/c$^2$ neutrino. So we do not need to worry about 
nonzero graviton mass for our problem and we consider in the following
that GW propagate at the speed of light $c$.

Neutrinos with a mass $m_\nu$ will arrive at the detectors with a 
propagation time delay $\Delta t_{prop}$ given by
\beqn\label{delay}
  \Delta t_{prop} &=& \frac {L}{2~c}~\left(
                      \frac {m_\nu c^2}{E_\nu}\right)^2 \\
             &\simeq& 5.15~{\rm ms}~\left(\frac {L}{10~{\rm kpc}}\right) 
                         ~\left(\frac {m_\nu c^2}{1~{\rm eV}}\right)^2 
                         ~\left(\frac {10~{\rm MeV}}{E_\nu}\right)^2~.
\eeqn
To this time should be added any time difference at the source 
and another propagation delay because the neutrino and gravitational
wave detectors are not located at the same site. The latter can only be
derived when the source direction is known which can be achieved if the
bursts are registered in coincidence by several detectors using triangulation.
Of course the most precise determination is expected to come from
optical telescopes, typically a few days later, when the supernova explosion
finally occurs. Finally it is assumed that the distance L will be
derived from the optical measurements but it should also be pointed out
that a reasonable estimate of the distance can be deduced from the
absolute event rate in the neutrino detectors. Indeed the total energy 
release in the collapse is directly related to the mass
of the iron core which can be reasonably estimated
with an uncertainty of typically $40~\%$~\cite{bethe} 
leading to a $20~\%$ accurate measurement of the distance.

It is interesting to consider the precision which can be obtained
on the quantity of interest, $\Delta t_{prop}$, hence on $m_\nu$. The
observed time difference between the $\nu_e$ flash and the maximum of
the gravitational waveform is
\beqn
\label{totdelay}
 \Delta t_{\nu,GW} ~=~ \Delta t_{prop} +  \Delta t_{\nu_e,bounce}
                     -  \Delta t_{GW peak,bounce}
\eeqn
with obvious notations. We examine in turn the two model-dependent 
terms to Equation~(\ref{totdelay}) already discussed earlier
and the two contributions to the experimental error on
$\Delta t_{\nu,GW}$:
\begin{itemize}
\item $\Delta t_{GW peak,bounce}$ is expected to be very small. The
value ($0.1 \pm 0.4$) ms is obtained from the library of waveforms 
produced in Ref.~\cite{zm} as shown in Fig.~\ref{zm_tbmax}. The various
entries correspond to different sets of parameters used in the
simulation. The initial angular momentum and the compressibility of the
supernuclear matter are important input variables in this respect.
The time range obtained thus represents a realistic coverage
of the core collapse parameters.
\item $\Delta t_{\nu_e,bounce}$ is discussed above with the
estimate ($3.5 \pm 0.5$) ms, where the error reflects the uncertainties
in the shock wave propagation.
\item the measurement of the GW timing depends on the signal-to-noise
ratio $\rho$ in the detector, itself a function of the detection algorithm 
used to filter out the signal corresponding to the GW burst. Previous 
studies of robust filters~\cite{lal1,lal2} provide a timing uncertainty 
~\cite{lal3} given by $\delta t_{GW}^{peak} \sim 1.45 \tau / \rho$,
where $\tau$ is the rms width of the main GW peak. For the signals
simulated in Ref.~\cite{zm} $\tau \sim 1$ ms and the mean value of
$\rho$ is very close to 10 for supernovae located at 10 kpc, 
yielding a GW timing uncertainty of 0.15 ms.
\item the determination of the mean timing of the $\nu_e$ flash depends
on the event statistics $N_\nu$ and the flash width $\sigma_{flash}$
through  $\delta t_\nu^{peak} = \sigma_{flash} / \sqrt{N_\nu}$,
scaling as $L$. This translates into an uncertainty on
$m_\nu^2$ independent of the supernova distance $L$, as
$\delta m_\nu^2 \propto \delta t / L$. Simulations~\cite{simul} 
indicate that $\sigma_{flash} \sim (2.3 \pm 0.3)$ ms.
\end{itemize}

The total timing uncertainty can therefore be cast into two components:
one of statistical nature, dominated by $\delta t_\nu^{peak}$, and the 
other originating from systematic sources, 
estimated from above to be 0.65 ms. It may
be possible to reduce this systematic uncertainty with the observation
of an actual supernova event, since additional measurements such as the
neutrino energy spectrum and the shape of the GW waveform can provide
constraints on the core collapse phenomenology within the framework
of existing simulation codes.
%
%
\section{Different Methods and Results}
Several methods taking into account the time correlation between GW and 
neutrino bursts can be envisaged, depending on the neutrino detector 
type and the $\nu_e$ survival probability.

\subsection{Method 1: $\nu_e$ detection in SNO}

{\it Method 1} relies on the detection of the $\nu_e$ flash in SNO
through reaction~(\ref{nudcc1}) providing good timing and energy information.
This approach is the best when the $\nu_e$ survival probability is large
enough (the precise value depends on the supernova distance). In this case 
the $\nu_e$ peak is well separated from the thermal distribution and its
timing should be easily determined given enough events, {\it i.e.} for
distances up to 13 kpc. 

\begin{figure}
\begin{center}
\epsfxsize 14cm
\epsffile{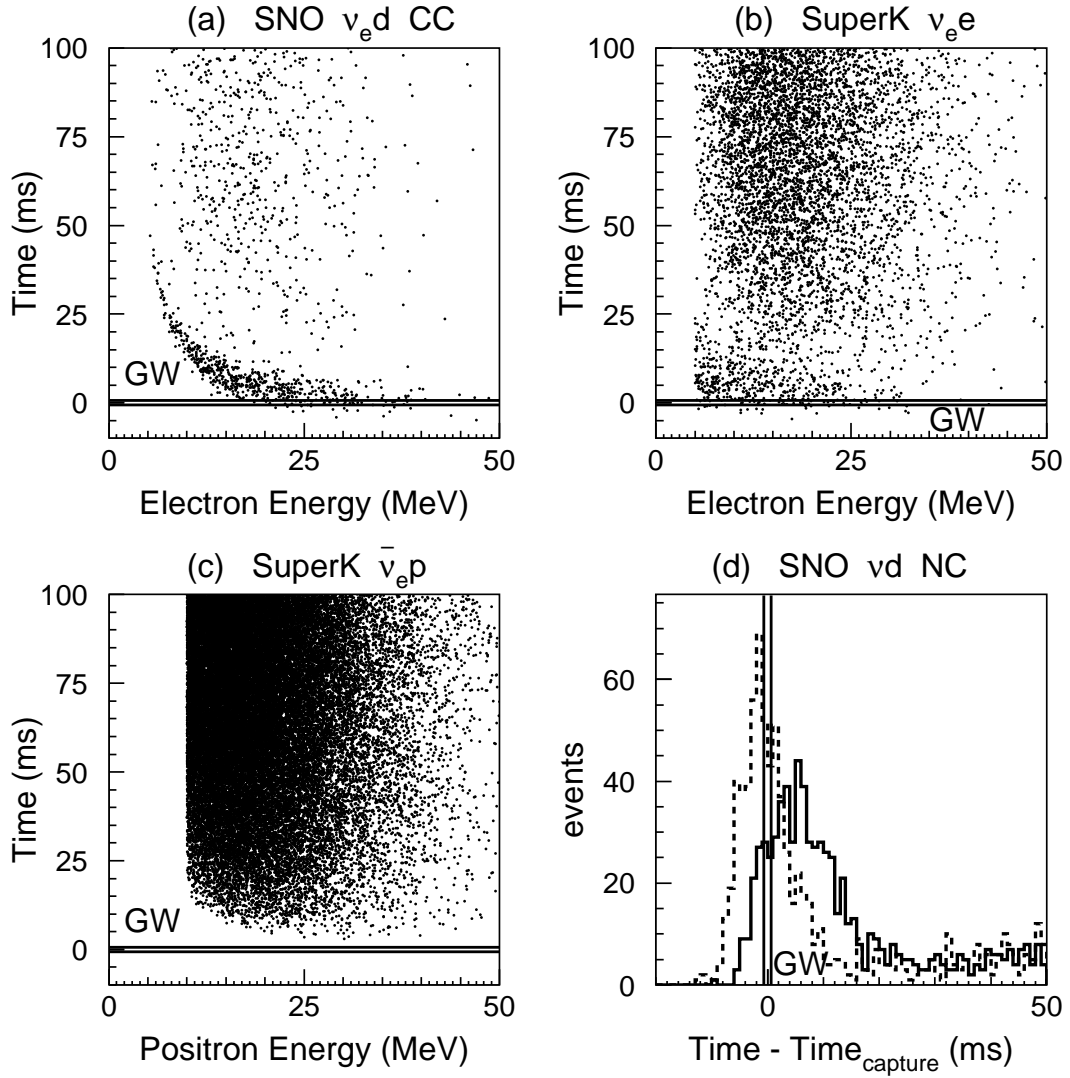}
\end{center}
\caption{\it Illustration of the four proposed methods with simulated data
             in neutrino detectors. Each situation corresponds to a
             supernova collapse at 10 kpc assuming 
             a neutrino mass of 2 eV/c$^2$.
             The expected statistics is scaled up by a factor of 100 in order
             to better visualize the distributions. In the first three cases
             events are displayed as function of electron (positron) energy
             and arrival time (defined such that zero-mass neutrinos arrive
             at t=0, as defined by the GW timing): (a) method (1) using the
             $\nu_e d$ process in SNO, including background from thermal
             $\nu_e$'s and $\overline{\nu}_e$'s; (b) method (2) based on
             $\nu_e e$ elastic scattering in SuperKamiokande, with background
             from  the $\overline{\nu}_e p$ process reduced by cuts;
             (c) method (3) using the $\overline{\nu}_e p$ reaction in
             SuperKamiokande. Finally in (d) the time distribution of
             neutral-current events is shown in SNO where the neutrino timing
             has been shifted by the average time for neutron capture 
             (the dashed histogram corresponds to a zero-mass neutrino).
             In all four plots the double line labelled GW indicates the
             $\pm~1~\sigma$ gravitational wave timing.} 
\label{simexp}
\end{figure}

We have performed simulations 
of supernova detections over a range of distances, using $P_e = 0.5$, 
the characteristics of the SNO detector~\cite{sno} 
and the estimate of timing accuracies given in the preceding section. 
An electron detection threshold of 5 MeV has been conservatively assumed:
whereas the present threshold used by SNO for solar neutrinos is only
6.75~MeV~\cite{sno_solnu}, the large instantaneous rate of a supernova
would allow one to lower the analysis threshold 
essentially down to the hardware value
of 2~MeV~\cite{jklein}. Neutrino energies are generated according to 
a Fermi-Dirac distribution with a characteristic temperature of 3.5 MeV.
The 2-dimensional distribution of relative arrival time and neutrino energy 
is displayed in Fig.~\ref{simexp} for the distance $L = 10$~kpc and a 
neutrino mass of 2~eV/c$^2$, but with a statistics 
enlarged by a factor of 100 in 
order to better visualize the problem. The neutronization peak is spread out 
with energy in a band which deviates from t=0 (the time delay between gravity
waves and zero-mass neutrinos has been subtracted out for clarity) in the 
lowest energy range. Although the assumed neutrino temperature corresponds to
a mean produced energy of 11~MeV, the observed average energy of the detected 
events is raised to 20~MeV because of the strong energy dependence 
of the cross section. A log likelihood fit of the event population
in the neutronization band yields the observed mass. 

It is clear from the
plot that, for small distances and consequently large neutrino rates,
SNO can determine the mass by itself if statistics is sufficient
to derive from the fit both the mass and the 'zero-mass' arrival time. This
is indeed the case since the Fermi-Dirac energy distribution of the neutrino
energy is wide enough to sample situations sensitive (low energy) or not 
(high energy) to the mass, while the situation is reversed for the 
determination of the prompt arrival time. However the approach without 
independent timing information deteriorates rapidly with increasing
distances as statistics at higher neutrino energies becomes insufficient to
pin down the zero-mass time. One therefore expects GW timing to become
increasingly helpful. 

This expectation is verified by the results of
2-dimensional maximum likelihood fits to 
data of many simulated experiments, as shown in Fig.~\ref{all_res}:
the total uncertainty $\delta m_\nu^2$ is found to be about 
0.5-0.6~eV$^2$/c$^4$, essentially independent of the distance up to 13~kpc
when statistics runs out, as expected. If no GW timing information is
available the accuracy steadily deteriorates with the distance, reaching 
1.5~eV$^2$/c$^4$ at 10~kpc and running out of events for a joint determination
of both neutrino mass and zero-mass arrival time.

\subsection{Method 2: $\nu_e$ detection in SuperKamiokande}

{\it Method 2} applies under the same conditions as for {\it Method 1}
with SNO, but this time using $\nu_e$ elastic scattering on electrons in
the SuperKamiokande detector. Given the relative masses of the detectors
and the relevant cross sections, it turns out that the statistics
is similar in both cases. An apparent disadvantage of this method is that
neutrino energy information is strongly reduced. However the
loss of information on the neutrino mass is not large as time delays are
preserved and a 2-dimensional likelihood fit still captures the
essential features of the experimental distribution for most of 
the distance range.
Background from reaction~(\ref{antinup}) must be suppressed:
a rejection factor of 10 is achieved through a cut, $\cos{\theta}>0.8$,
where $\theta$ is the angle between the electron and the supernova
directions. The latter is assumed to be known from
optical observations later on.

Using an electron detection threshold of 5~MeV, likelihood
fits of simulated data such as shown in Fig.~\ref{simexp} are performed,
yielding the sensitivity curve in Fig.~\ref{all_res} with
GW timing. As expected the value for $\delta m_\nu^2$ is similar to that
obtained in {\it Method 1}, $\sim 0.5-0.7 ~{\rm eV}^2$/c$^4$. 
As observed in {\it Method 1},
it is still possible to fit the distribution without an a priori
knowledge of the absolute timing provided by GW detection, but the
sensitivity is strongly reduced in this case.

\subsection{Method 3: $\overline{\nu}_e$ detection}

{\it Method 3} relies on the onset of $\overline{\nu}_e$
thermal production, detected essentially through the more copious 
reaction~(\ref{antinup}) used by essentially all neutrino experiments.
For a supernova at 10 kpc the expected rates are 5300, 400, 135, 
and 133, in SuperKamiokande, SNO, LVD, and MACRO, respectively.
Contrary to the neutronization flash the time distribution 
of the thermal $\overline{\nu}_e$'s is more model-dependent. 
With a characteristic risetime is $\sim 50$ ms,
the shape of the spectrum on the time scale of 1-10 ms is hard to
control theoretically, but the onset is closely related to the timing
of the neutronization flash~\cite{mazurek,burrows_kg}. This feature is
supported by extensive simulation work~\cite{simul}. 

The advantages of this
method is the availability of the neutrino energy measurement and the fact
that the rate is essentially insensitive to neutrino oscillations. A cut 
on the positron angle with respect to the supernova direction,
$\cos{\theta} < 0.8$, has to be applied in order to remove the few 
forward-peaked electron events from elastic scattering 
with a signal loss of only $10~\%$. The positron energy threshold
has to be raised to 10~MeV in order to avoid the background of
$\gamma$-rays from nuclear de-excitation induced by the neutral-current
neutrino processes~\cite{langanke}. This method can be implemented 
without GW information~\cite{totani}, but the sensitivity is
greatly enhanced by GW timing. A simulated distribution is given
in Fig.~\ref{simexp}(c) and likelihood fits yield the precision shown in
Fig.~\ref{all_res}, with typically $\delta m_\nu^2 \sim 0.9 ~{\rm eV}^2$/c$^4$.

Although all other methods are limited to a supernova distance of 13~kpc
because of neutrino statistics, {\it Method 3} 
does not suffer from this limitation.
The rate expected in SuperKamiokande would still be sufficient up to
$\sim 100$~kpc, however the sensitivity of present GW detectors is such that
one can hardly consider detections beyond our galaxy~\cite{lal1,lal2,lal3}.

\subsection{Method 4: $\nu_{\mu,\tau}$ detection in SNO}

Finally, {\it Method 4} needs to be used if neutrino oscillations
turn the $\nu_e$'s in the neutronization peak into $\nu_{\mu,\tau}$'s. Again
SNO is the only neutrino experiment able to exploit this possibility. The
situation is however much less favourable than 
in {\it Method 1}, as {\it (i)} the neutral-current 
cross sections are a factor 2.5 smaller than their 
charged-current counterparts in the 10-40 MeV range, {\it (ii)} the neutrino 
energy information is lost, and {\it (iii)} timing is degraded by the
fluctuations in the neutron capture. In this case absolute timing from
GW detection is crucial whatever the supernova distance. The fact that 
no energy information is available means a greater dependence on the
model for the shape of the neutronization peak. There is also some 
uncertainty on the mean capture time, but it can be experimentally 
calibrated using reaction~(\ref{nudcc2}) which provides signals 
from both the prompt positron and the delayed neutron capture 
with good statistics.

\begin{figure}
\begin{center}
\epsfxsize 12cm
\epsffile{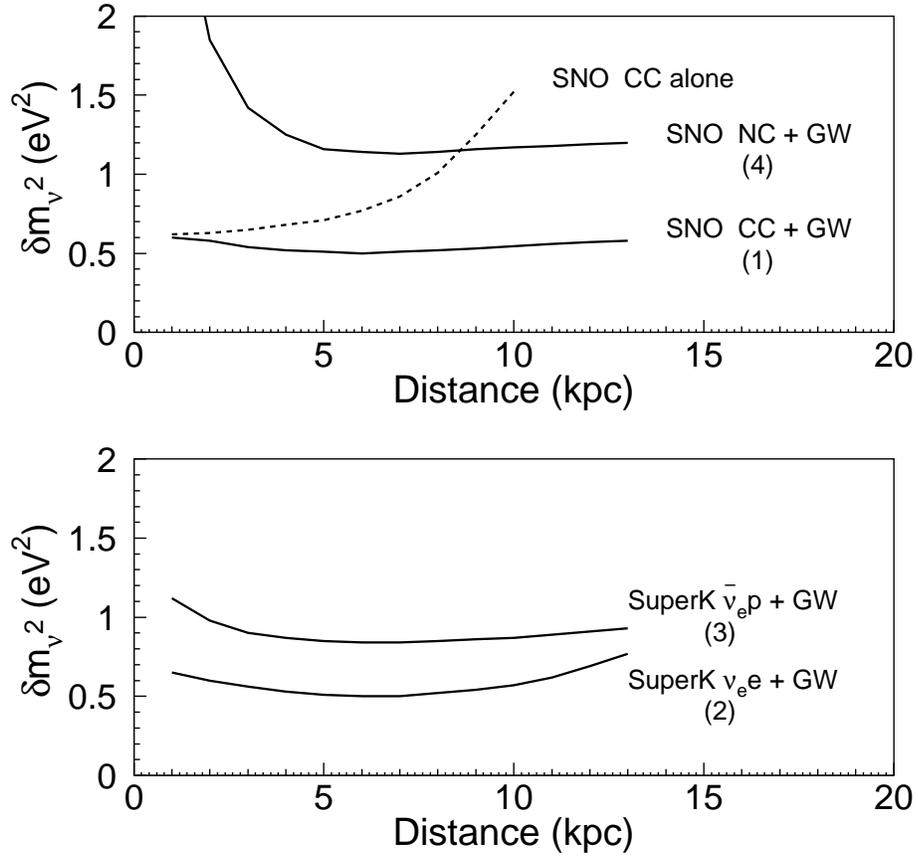}
\end{center}
\caption{\it The estimated sensitivity $\delta m_\nu^2$ of Methods 
             1 to 4 with the SNO (a) and SuperK (b) detectors, for
             stellar collapses as function of distance. Results for Methods
             1,2 (resp. 4) are given for $P_e=0.5$ (resp. $P_e \sim 0$),
             while Method 3 applies independently of $P_e$.}
\label{all_res}
\end{figure}

Examples of simulated time distributions 
are given in Fig.~\ref{simexp}(d). Assuming full 
$\nu_e$-to-$\nu_{\mu,\tau}$ conversion, the estimated uncertainty of this
method is found to be $\delta m_\nu^2 \sim 1.2 ~{\rm eV}^2$/c$^4$ from
a fit of the neutrino time distribution with respect to the GW signal.
The sensitivity, shown in Fig.~\ref{all_res}, is 
strongly degraded at small distances where the systematic 
uncertainty  on the relative timing dominates over
the statistical error of neutrino timing. As expected this method is less
sensitive than {\it Method 1}, but it is the only choice left 
if the $\nu_e$ survival probability is too small.

The sensitivities expected with the different methods are summarized in
Table~\ref{table_results} for the two scenarios of large ($P_e = 0.5$) 
and of negligible $\nu_e$ survival probabilities. They depend rather weakly 
on the supernova distance and they are given at 10 kpc. Since the neutrino
statistics are uncorrelated between the different methods, the overall
sensitivity using all four approaches can be correspondingly improved to
$\delta m_{\nu_e}^2 \sim 0.35 ~{\rm eV}^2$/c$^4$ for $P_e = 0.5$ and 
$\sim 0.7 ~{\rm eV}^2$/c$^4$ for $P_e \sim 0$. In case the experiments 
do not see any deviation from nonzero mass $95\%$ C.L. upper limits 
of 0.75 and 1.1 eV/c$^2$ will be derived on the degenerate neutrino mass 
in the two scenarios, respectively.

\vspace{2cm}

\begin{table}
\setlength{\tabcolsep}{1pc}
\begin{center}
{\normalsize
\begin{tabular}{ccc} 
method    & $P_e = 0.5$ & $P_e \sim 0$  \\ \hline
 1      & $0.55$ &   -    \\
 2      & $0.57$ &   -    \\
 3      & $0.87$ & $0.87$ \\
 4      & $1.63$ & $1.15$ \\ \hline
combined& $0.35$ & $0.69$ \\ 
\end{tabular}
}
\caption[.]{\label{table_results}\it
            Total uncertainties $\delta m_\nu^2$ (eV$^2$/c$^4$) expected 
            in the four proposed methods under two scenarios for neutrino 
            oscillations. The values are quoted for a supernova at 10 kpc,
            but they are weakly dependent on the distance. See details 
            in the text.}

\end{center}
\end{table}
%
%
\section{Conclusions and Prospects}
The next type-II supernova explosion in the Galaxy is expected to provide
extremely valuable information on neutrino masses. New methods, based on 
the availability of massive neutrino detectors and the near-operation of new 
large interferometric gravity-wave antennas, have been proposed. They rely 
on the time coincidence between neutrino and gravitational wave 
detections. Different experimental approaches have to be considered depending 
on the capabilities of the various neutrino detectors and on the overall 
effect of oscillations between the three neutrino flavours. 
 
The most sensitive method is based on the detection by SNO and 
SuperKamiokande of prompt electron neutrinos from the neutronization 
peak which is tightly correlated in time with the bounce 
terminating the stellar core collapse, itself corresponding to 
the maximum gravity wave activity. If the $\nu_e$ survival probability is
large, this method yields a $\nu_e$ mass 
sensitivity for each detector almost independent of the supernova 
distance up to 13~kpc, measured by 
$\delta m_\nu^2 \sim 0.60 ~{\rm eV}^2$/c$^4$. 
The combination of the results from SNO and SuperKamiokande would directly
exclude a $\nu_e$ mass of 0.75~eV/c$^2$ at $95~\%$ C.L. if no significant mass 
effect were found. This value is a factor of 4 smaller than current limits 
from end-point tritium experiments. If the mass were indeed 2~eV/c$^2$ 
the expected effect would correspond to a 11 $\sigma$ deviation 
from zero-mass and the mass would be measured with a precision of $4.5~\%$. 
A 1~eV/c$^2$ $\nu_e$ mass would still be seen at the 3~$\sigma$ level
and determined with a precision of  $17~\%$. 
Two specific methods are proposed if neutrino conversions in the outer 
star mantle disfavours $\nu_e$ detection. One still uses the neutronization 
peak and neutral-current detection in SNO, while the other is based on 
the measurement of the onset of thermal $\overline{\nu}_e$ production 
in SuperKamiokande. When combined they still provide a sensitivity of
$\delta m_\nu^2 \sim 0.7 ~{\rm eV}^2$/c$^4$ and a $95~\%$ C.L upper limit
of 1.1 eV. Finally it is interesting to note that these results 
follow from time differences accurately measured at a level of 
$\sim 10^{-15}$ of the total time-of-flight. 

The approach can be extrapolated to the next generation of neutrino and
gravitational wave detectors. A valuable goal would be to bridge the gap
between the reachable mass value with present detectors (0.7~eV/c$^2$) and the
upper range provided by neutrino oscillations in the least-degenerate
neutrino mass scenario (0.06~eV/c$^2$). This requires a factor of 100 
increase in the neutrino detector masses (HyperKamiokande?), which 
would be matched to the factor of 10 improvement in sensitivity 
considered for GW antennas on the timescale of 6-7 years~\cite{ligo2}.
Such a desirable situation would have a number of advantages:
{\it (i)} the precision on the neutrino timing for a supernova 
detection would be improved by a factor of 10, {\it (ii)} distances 
up to 150~kpc could be reached with the proposed methods 
with a corresponding gain in the supernova rate, 
and {\it (iii)} the large statistics that would
be available for a galactic event would permit a much better understanding
of the collapse dynamics, hence offering the possibility to better
control the systematic timing uncertainty from the models.
%
%

%
%
{
}

\end{document}